\documentclass{emulateapj}
\usepackage{apjfonts}
\usepackage[]{natbib}
\usepackage{graphics}
     
\newcommand{\etal}{et al.}  
\newcommand{\per}{\ensuremath{^{-1}}}

\newcommand{\hal}{H\ensuremath{\alpha}}
\newcommand{\hbeta}{H\ensuremath{\beta}} 

\newcommand{\msun}{\ensuremath{M_{\odot}}}
\newcommand{\lsun}{\ensuremath{L_{\odot}}}
\newcommand{\kms}{km s\ensuremath{^{-1}}}

\newcommand{\mbh}{\ensuremath{M_\mathrm{BH}}}

\newcommand{\sigmastar}{\ensuremath{\sigma_\star}}

\newcommand{\msigma}{\ensuremath{\mbh-\sigma}}

\newcommand{\rosat}{\emph{ROSAT}}
\newcommand{\lbol}{\ensuremath{L_{\mathrm{bol}}}}

\newcommand{\ledd}{\ensuremath{L_{\mathrm{Edd}}}}
\newcommand{\chandra}{\emph{Chandra}}

\slugcomment{to appear in The Astrophysical Journal Letters}
\shorttitle{OFFSET AGN IN NGC 3341} 
\shortauthors{BARTH ET AL.}

\begin{document} 

\title{An Offset Seyfert 2 Nucleus in the Minor Merger System NGC 3341}

\author{Aaron J. Barth\altaffilmark{1}, Misty
  C. Bentz\altaffilmark{1}, Jenny E. Greene\altaffilmark{2,3}, and Luis
  C. Ho\altaffilmark{4}}

\altaffiltext{1}{Department of Physics and Astronomy, 4129 Frederick
  Reines Hall, University of California, Irvine, CA 92697-4575;
  barth@uci.edu}

\altaffiltext{2}{Department of Astrophysical Sciences, Princeton
  University, Princeton, NJ 08544}

\altaffiltext{3}{Hubble Fellow and Princeton-Carnegie Fellow}

\altaffiltext{4}{The Observatories of the Carnegie Institution of
  Washington, 813 Santa Barbara Street, Pasadena, CA 91101}

\begin{abstract}

We present the discovery of a triplet of emission-line nuclei in the
disturbed disk galaxy NGC 3341, based on archival data from the Sloan
Digital Sky Survey and new observations from the Keck Observatory.
This galaxy contains two offset nuclei within or projected against its
disk, at projected distances of 5.1 and 8.4 kpc from its primary
nucleus and at radial velocity separation of less than 200 \kms\ from
the primary. These appear to be either dwarf ellipticals or the bulges
of low-mass spirals whose disks have already been stripped off while
merging into the primary galaxy.  The inner offset nucleus has a
Seyfert 2 spectrum and a stellar velocity dispersion of $70\pm7$ \kms.
The outer offset nucleus has very weak emission lines consistent with
a LINER classification, and the primary nucleus has an emission-line
spectrum close to the boundary between LINER/\ion{H}{2} composite
systems and \ion{H}{2} nuclei; both may contain accreting massive
black holes, but the optical classifications alone are ambiguous.  The
detection of an offset active nucleus in NGC 3341 provides a strong
suggestion that black hole accretion episodes during minor mergers can
be triggered in the nuclei of dwarf secondary galaxies as well as in
the primary.

\end{abstract}


\section{Introduction}

In the standard hierarchical merger paradigm for galaxy formation and
evolution, a natural consequence of the merger of two galaxies
containing massive black holes is that the two black holes will sink
to the center of the host galaxy by dynamical friction, eventually
forming a binary and merging together (see Merritt \& Milosavljevi\'c
2005 for a review).  As a result, there is a general expectation that
some merging and post-merger galaxies should have two massive black
holes.  However, definitive observational identification of galaxies
containing two (or more) massive black holes is limited to objects
showing evidence for accretion-powered activity from spatially
resolved active galactic nuclei (AGNs) \citep{kom03a}.  A galaxy with
two widely separated active nuclei is often referred to as a ``dual
AGN'' to distinguish it from the case of a ``binary AGN'' in which the
two black holes form a tight, gravitationally bound binary.  Several
examples of dual AGNs are now known
\citep[e.g.,][]{owe85,kom03b,bal04,gua05,ger07,bia08}, most of which
are found in major mergers of massive galaxies.  It is also expected
that minor mergers can trigger episodes of nuclear activity
\citep{hm95}, and dual AGNs could also be present in some minor
mergers with dwarf secondary galaxies if the secondary galaxies also
harbor central black holes.  However, the fraction of dwarf or
low-mass galaxies that contain black holes is unknown, and only weakly
constrained by AGN surveys \citep{gh07} and by stellar-dynamical
searches for black holes in dwarf galaxies \citep{val05}.

In this \emph{Letter}, we present the discovery of a triplet of
emission-line nuclei (at several kpc separation) in NGC 3341.  This
system appears to contain two dwarf galaxies merging with a massive
disk galaxy, and one of the offset nuclei has a Seyfert 2 spectrum.
NGC 3341 is therefore a rare example of a disk galaxy with an offset
AGN, and an excellent nearby case study for investigations of AGN
triggering during a minor merger.  The primary nucleus and the second
offset nucleus have optical spectra that indicate possible low-level
AGN activity as well, making this a candidate dual or even triple
AGN system.

\section{Archival SDSS Data}

In the course of searching the Sloan Digital Sky Survey (SDSS)
archives for dwarf galaxies with Seyfert-type spectra, we came across
the object SDSS J104232.05+050241.9.  Its SDSS spectrum is that of a
Seyfert 2 nucleus, with strong narrow emission lines superposed on a
starlight-dominated continuum.  Its $g$-band apparent magnitude is
18.30 mag; at $z=0.0271$ (or $D_L\approx117$ Mpc for $H_0=71$ km
s\per\ Mpc\per, $\Omega_M = 0.27$, and $\Omega_\Lambda=0.73$), this
corresponds to only $M_g = -17$ mag.  In the online MPA/JHU
catalog\footnote{http://www.mpa-garching.mpg.de/SDSS} of AGNs in SDSS
Data Release 4 \citep{kau03}, it is listed as an AGN with a host
galaxy stellar mass of $1.6\times10^9$ \msun\ and an [\ion{O}{3}]
$\lambda5007$ luminosity of $2\times10^6$ \lsun.  This object
initially appeared to be an excellent example of an AGN in a dwarf
host galaxy, but when we examined the SDSS image of the object, we
found that it is not an individual dwarf galaxy.  Instead, it is an
off-nuclear knot within the disk of the larger galaxy NGC 3341.  The
SDSS catalog magnitude of $g=18.30$ mag corresponds only to the
brightness of this off-nuclear source, while NGC 3341 overall is a
giant disk galaxy with $B=14.9$ mag (or $M_B=-20.3$ mag).

\begin{figure}
\plotone{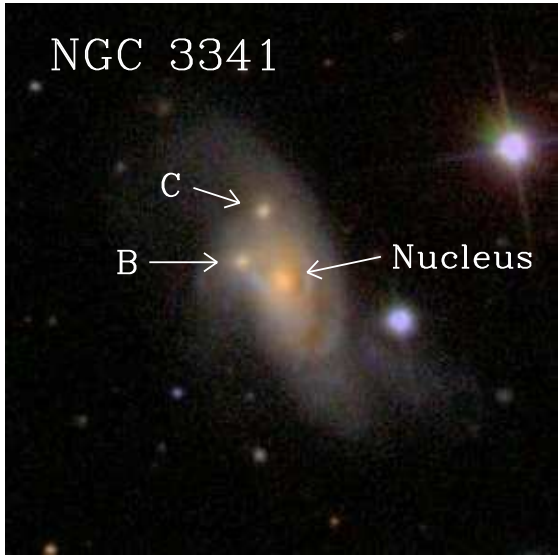}
\caption{SDSS color-composite image of NGC 3341.  The field of view is
  $2\arcmin\times2\arcmin$.  North is up and east is to the left.
\label{sdssimage}}
\end{figure}

Figure \ref{sdssimage} shows the SDSS image of NGC 3341.  It is a
disturbed disk galaxy (Hubble type $\sim$Sa-Sb, but listed in the UGC
catalog as ``peculiar'') undergoing a minor merger, with two dwarf
companions within (or projected against) the galaxy disk.  We label
these off-nuclear sources as NGC 3341B and NGC 3341C, with object B
corresponding to SDSS J104232.05+050241.9.  The SDSS only obtained a
spectrum of object B, and not of the galaxy's primary nucleus or of
object C.  The luminosities of the two off-nuclear sources ($M_g =
-17$ and $-16.6$ mag for objects B and C) are too high for these to be
super star clusters, but they could be dwarf ellipticals, or the
bulges of low-mass spirals whose disks have already been stripped off
during the merger.  Objects B and C are located at projected
separations of 9\farcs5 and 15\farcs6 (5.1 and 8.4 kpc) from the
primary nucleus, respectively, and the separation between B and C is
11\farcs9 (6.4 kpc).

\section{Keck Observations}

\begin{figure}
\plotone{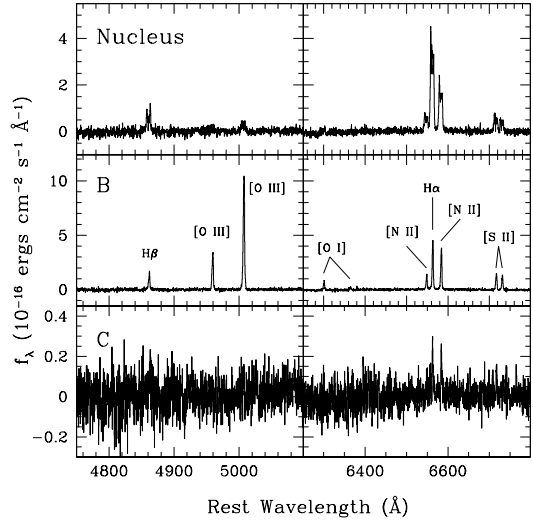}
\caption{Portions of the starlight-subtracted Keck spectra of the
three nuclei.
\label{spectra}}
\end{figure}

We obtained spectra of all three nuclei at the Keck-II telescope using
the ESI spectrograph \citep{she02} on the night of 2008 March 2 UT.
Observing conditions were relatively poor for Mauna Kea, with some
cirrus overhead and seeing of $\sim1\farcs5$.  Despite the poor
seeing, a narrow slit of width 0\farcs75 was used in order to resolve
narrow absorption and emission lines in dwarf galaxies.  The spectra
cover 3800--10900 \AA\ over 10 echelle orders, at a uniform scale of
11.5 km s\per\ pixel\per\ and instrumental dispersion of $\sigma_i=22$
\kms.  The spectra were obtained with the slit oriented at
PA=115\arcdeg, close to the parallactic angle for the midpoint of the
exposure sequence, at airmass 1.2--1.3.  Exposure times were 900 s
each for the primary nucleus and object C, and 1200 s for object B.
Spectra were extracted using a width of 2\arcsec\ and flux-calibrated
using observations of the standard star Feige 34.  The $S/N$ per pixel
in the continuum at 6600 \AA\ is $\sim20$, 15, and 10 for the nucleus
and objects B and C, respectively.  To obtain pure emission-line
spectra, we performed starlight subtraction using observations of
template stars observed with the same ESI configuration.  (See Barth
\etal\ 2008 for a detailed description of the reduction, calibration,
and starlight-subtraction procedures.)  The starlight-subtracted
spectra of the three nuclei are shown in Figure \ref{spectra}.  For
object C, the only detected emission lines are \hal\ and [\ion{N}{2}]
$\lambda6583$.

\section{Results and Discussion}

The starlight subtraction procedure yields radial velocities of $cz =
8325\pm9$, $8131\pm6$, and $8145\pm6$ \kms\ for the primary nucleus
and objects B and C, respectively.  This confirms the physical
association between the three objects and rules out a chance
projection of unrelated sources.  

\begin{figure}
\plotone{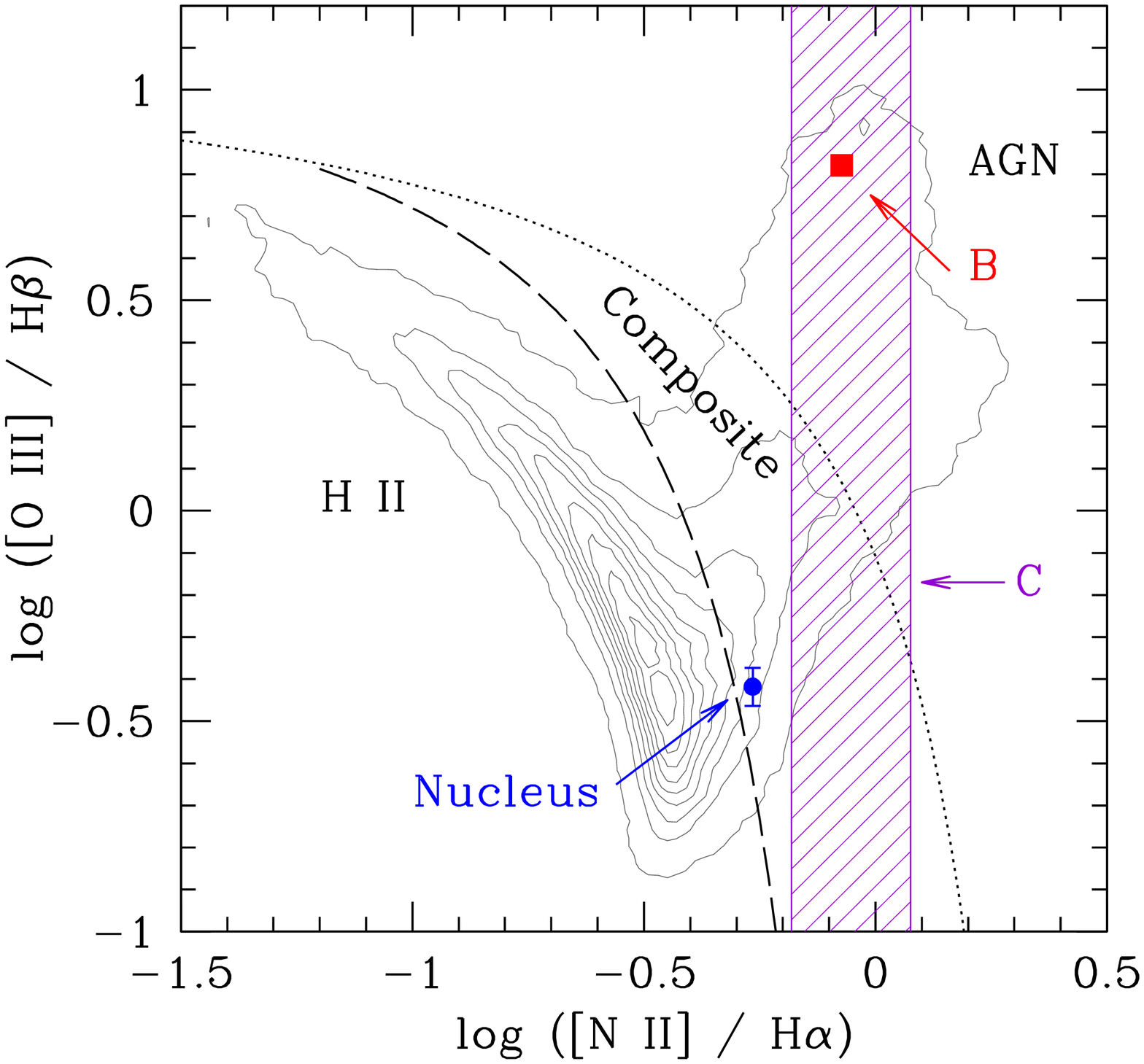}
\caption{Diagnostic diagram showing the emission-line ratios of the
  three nuclei.  Black contours denote emission-line galaxies from the
  \citet{kau03} sample.  The dashed and dotted curves represent the
  classification criteria of \citet{kew06}, in which \ion{H}{2} nuclei
  fall below and to the left of the dashed curve, AGNs lie above and
  to the right of the dotted curve, and composite objects lie in the
  intermediate region.  The primary nucleus of NGC 3341 is shown as a
  blue circle, and object B as a red square.  For object B and for
  the [\ion{N}{2}]/\hal\ ratio at the primary nucleus, the error bars
  are smaller than the plot symbols.  The violet shaded band gives the
  $1\sigma$ uncertainty range for the [\ion{N}{2}]/\hal\ ratio of
  object C.
\label{bpt}}
\end{figure}

Emission-line fluxes were measured by direct integration of the
starlight-subtracted spectra; the key diagnostic flux ratios are
listed in Table \ref{table1}.   Figure \ref{bpt} displays the
emission-line flux ratios of [\ion{O}{3}]/\hbeta\ and
[\ion{N}{2}]/\hal\ in a \citet{bpt81} diagnostic diagram, in
comparison with galaxies from the \citet{kau03} sample.  The Keck
spectrum of object B confirms its classification as a Seyfert 2, as it
falls within the Seyfert branch of the diagnostic diagram.  Its
[\ion{O}{3}] $\lambda5007$ emission line has FWHM = 116 \kms, similar
to the low [\ion{O}{3}] linewidths seen in other Seyfert 2 nuclei
having low-mass host galaxies \citep{bgh08}.

In the [\ion{O}{3}]/\hbeta\ vs.\ [\ion{N}{2}]/\hal\ diagram, the
primary nucleus falls in the region of LINER/\ion{H}{2} ``composite''
or ``transition'' objects, close to the boundary with \ion{H}{2}
nuclei, following the \citet{kew06} classification scheme.  However,
we note that the [\ion{O}{1}]/\hal\ and [\ion{S}{2}]/\hal\ ratios at
the nucleus are consistent with classification as an \ion{H}{2}
nucleus \citep{hfs97, kew06}.  Thus, the classification of the nucleus
is somewhat ambiguous.  The narrow-line profiles at the nucleus are
double-peaked, with a peak-to-peak velocity separation of 310 km
s\per.  This velocity splitting could arise from either radial motion
or rotation; similar double-peaked profiles due to the rotation of a
circumnuclear disk are seen, for example, in the transition-type
nucleus of NGC 3245 \citep{hfs95}.

For the weak emission lines in object C, we find [\ion{N}{2}]/\hal\
$=0.93 \pm 0.27$, indicated as a shaded band in Figure \ref{bpt}.  The
nearly-equal strengths of [\ion{N}{2}] and \hal\ are inconsistent with
an \ion{H}{2} region origin for the emission lines.  Given the
non-detection of [\ion{O}{3}] $\lambda5007$ emission, the most likely
classification for this object is a LINER or LINER/\ion{H}{2}
composite.  The \hal\ equivalent width is only $\sim1$ \AA.

While the optical data identify object B as an AGN with a high degree
of confidence, the source of ionizing photons in the nucleus and
object C is less certain.  In the absence of definitive AGN signatures
such as broad-line emission, very weak LINER-type emission lines such
as those seen in object C could in principle result from
photoionization of diffuse gas by sources unrelated to an AGN, such as
post-AGB stars \citep{bin94} or X-ray binaries.  Indeed, \chandra\
observations have revealed that some nearby ``Type 2'' LINERs do not
have a central AGN-like point source at all, but do have X-ray
emission dominated by a circumnuclear population of X-ray binaries
\citep{era02, flo06}.  As for the primary nucleus, the source of
ionization in LINER/\ion{H}{2} transition nuclei has been the subject
of much debate \citep[see][]{ho08}.  A likely interpretation is that
they are composite systems containing both a weak AGN and \ion{H}{2}
regions ionized by hot stars \citep[e.g.,][]{hfs93, vgv97, kew06}.
Shock heating could play some role in the narrow-line region
excitation as well \citep[e.g.,][]{ds95, cv01}.  Summarizing results
from X-ray surveys of nearby galaxies, \citet{ho08} concludes that the
majority of transition nuclei (perhaps $\sim75\%$) do contain AGNs.
Thus, the optical spectra of the nucleus and object C can be
interpreted as possibly containing low-luminosity accretion-powered
AGNs, but the optical data do not provide conclusive evidence for the
presence of AGNs in these objects.  A search for X-ray sources in NGC
3341 would potentially provide the best constraints on the AGN content
in this galaxy.

\begin{deluxetable}{lcccc}
\tablewidth{3in} \tablecaption{Emission-Line Ratios}
\tablehead{ \colhead{Object} & \colhead{[\ion{O}{3}]/\hbeta} & 
\colhead{[\ion{O}{1}]/\hal} &
\colhead{[\ion{N}{2}]/\hal} & \colhead{[\ion{S}{2}]/\hal}  }
\startdata
Nucleus & $0.38\pm0.04$ & $0.05\pm0.01$ & $0.54\pm0.02$ & $0.31\pm0.01$ \\
B       & $6.6\pm0.2$   & $0.20\pm0.01$ & $0.85\pm0.02$ & $0.66\pm0.02$ \\
C       & \nodata       & $<0.25$       & $0.93\pm0.27$ & $<0.6$ 
\enddata
\tablecomments{Forbidden-line measurements refer to [\ion{O}{3}]
$\lambda5007$, [\ion{O}{1}] $\lambda6300$, [\ion{N}{2}] $\lambda6583$,
and [\ion{S}{2}] $\lambda\lambda6716,6731$. }

\label{table1}
\end{deluxetable}

NGC 3341 is not detected in the \rosat\ All-Sky Survey, and no source
corresponding to its position is listed in the \rosat\ Faint Source
Catalog \citep{vog00}.  We estimate an upper limit of 0.02 counts
s\per\ in the \rosat\ image, corresponding to an upper limit of
$L$(0.1--2 keV)$<5\times10^{41}$ ergs s\per\ for an unabsorbed source
with a photon index of $\Gamma = 1.8$.  Using typical ratios of X-ray
to optical emission-line flux for nearby low-luminosity AGNs
\citep{ho08}, the expected luminosity of each of the three nuclei
would be at least an order of magnitude below the \emph{ROSAT} upper
limit.  Thus, much deeper observations would be needed in order to
search for X-ray sources in this system.  The primary nucleus is a
radio source, detected in the FIRST survey (Becker \etal\ 1995) with a
flux density of 4.8 mJy and deconvolved major and minor axis sizes of
$5\farcs5\times5\farcs0$.

We measured the stellar velocity dispersions of the three nuclei
following the direct-fitting methods described by \citet{bhs02}, by
fitting the spectra over the restricted wavelength range 5020--5500
\AA\ containing the strong Mg$b$ and Fe5270 features.  We find
$\sigmastar=182\pm14$, $70\pm7$, and $44\pm6$ \kms\ for the nucleus
and objects B and C, respectively.  If these objects follow the
\msigma\ correlation for quiescent galaxies of \citet{tre02}, the
expected black hole masses would be $9\times10^7$, $2\times10^6$, and
$3\times10^5$ \msun\ in the three nuclei.  \citet{hec04} show that a
typical [\ion{O}{3}] bolometric correction for Seyfert galaxies is
$\lbol/L$([\ion{O}{3}])$\approx 3500$ (with a scatter of approximately
0.38 dex).  With this bolometric correction, object B would have
$\lbol/\ledd \approx 0.1$.

From the 2MASS \citep{skr06} point source and extended source
catalogs, the $K_s$-band magnitudes of the primary galaxy and objects
B and C are 10.78, 14.26, and 15.07 mag, respectively.  To estimate
the stellar mass ratio of the merger, we use $K$-band mass-to-light
ratios from \citet{bell03}, computed using reddening-corrected optical
colors from SDSS.  Following the \citet{bell03} prescriptions, the
primary galaxy and objects B and C each have $M/L_K\approx0.85$.  This
yields a stellar mass of $\sim1\times10^{11}$ \msun\ for the primary
galaxy and a stellar mass ratio of 50:2:1 for the merging system
(primary:B:C).  We note, however, that this mass ratio does not
account for any stellar mass already stripped off of objects B and C
during the merger.  For a more detailed examination of the structure
of the merging system, higher-resolution imaging would be needed.
\emph{Hubble Space Telescope} data could clarify whether the offset
nuclei were originally spheroidal or elliptical galaxies, or the
bulges or pseudobulges of low-mass disk galaxies.

It is particularly interesting to find nuclear activity in at least
one of the secondary nuclei in this system, since the AGN fraction in
low-mass galaxies is known to be very small overall \citep{kau03,
gh07}.  Furthermore, the black hole occupation fraction in dwarf
galaxies is expected to be well below unity if the efficiency for
black hole seed formation is small in low-mass halos at high redshift
\citep{vol08}, or if gravity-wave kicks during binary black hole
coalescence can efficiently expel black holes from dwarf galaxies
\citep{mer04}.  There are examples of AGNs with host galaxy
luminosities similar to objects B and C, such as POX 52 \citep{bar04},
but they are predominantly isolated systems not involved in strong
interactions.  Simulations have shown that minor mergers can drive
radial inflows of gas to the center of the primary galaxy, potentially
triggering AGN accretion episodes \citep{hm95}, but there is little if
any statistical evidence for enhanced AGN activity in the secondary
galaxies of minor interaction systems.  In a study of close
(separation $<50h^{-1}$ kpc) galaxy pairs in SDSS, \citet{wg07} found
that in minor interactions with $\Delta m >2$ mag, there was a strong
increase in the AGN fraction for the primary galaxies (relative to a
matched field galaxy control sample) but there was no detectable
enhancement in the AGN fraction of the secondary galaxies in the pairs
relative to a matched control sample.  In view of the properties of
NGC 3341, we speculate that any enhancement in the AGN fueling rate in
the secondary might occur at very small separations ($\lesssim10$ kpc)
that are difficult to probe in statistical surveys.

This unusual AGN host galaxy was discovered because the SDSS
fortuitously selected object B as the one spectroscopic target in this
field.  If the primary nucleus had been targeted instead, the galaxy
would have appeared unremarkable, and follow-up observations of
objects B and C would have been unlikely.  What are the prospects for
detection of offset AGNs in other nearby galaxies?  \citet{kd05}
argued that some of the most extreme ultraluminous X-ray (ULX) sources
are likely to be the nuclei of captured satellite galaxies undergoing
AGN accretion episodes, rather than being X-ray binaries with
intermediate-mass black holes.  While no ULX sources are
currently known to have an origin in captured satellites, the
discovery of NGC 3341B in the SDSS suggests that optical selection can
be at least as efficient as X-ray selection for discovery of
off-nuclear, accreting black holes from satellite galaxies in minor
mergers.  However, optical selection depends crucially on the ability to
identify the nucleus of the secondary galaxy in imaging data.  As
discussed by \citet{wgb06}, there are severe observational biases that
make it very difficult to identify satellite galaxies involved in
minor interactions ($\Delta m > 2$ mag) at very close separations of
$\lesssim10$ kpc from the primary galaxy.  Furthermore, at separations
of a few kpc, satellite galaxies may be so severely disrupted that
they no longer have identifiable nuclei, and offset, low-luminosity
AGNs not embedded in stellar nuclei would likely be inconspicuous in
optical surveys.  Nevertheless, further searches for systems like NGC
3341 containing low-mass companions at small separations could provide
new constraints on the demographics of dual AGNs and the efficiency of
AGN fueling in minor mergers.

\acknowledgments

Research by A.J.B. and M.C.B. is supported by NSF grant AST-0548198.
Data presented herein were obtained at the W.M. Keck Observatory,
which is operated as a scientific partnership among Caltech, the
University of California, and NASA. The Observatory was made possible
by the generous financial support of the W.M. Keck Foundation.  The
authors wish to recognize and acknowledge the very significant
cultural role and reverence that the summit of Mauna Kea has always
had within the indigenous Hawaiian community.  Funding for the SDSS
and SDSS-II has been provided by the Alfred P. Sloan Foundation, the
Participating Institutions, the National Science Foundation, the
U.S. Department of Energy, the National Aeronautics and Space
Administration, the Japanese Monbukagakusho, the Max Planck Society,
and the Higher Education Funding Council for England.  This
publication makes use of data products from the Two Micron All Sky
Survey, which is a joint project of the University of Massachusetts
and the Infrared Processing and Analysis Center/California Institute
of Technology, funded by NASA and the NSF.

\end{document}